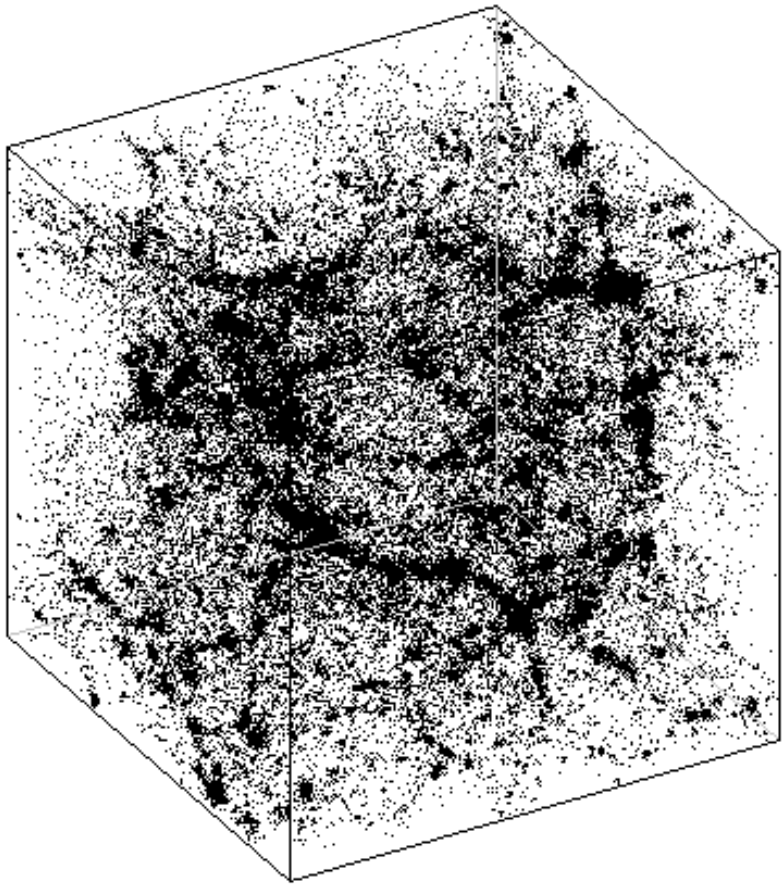

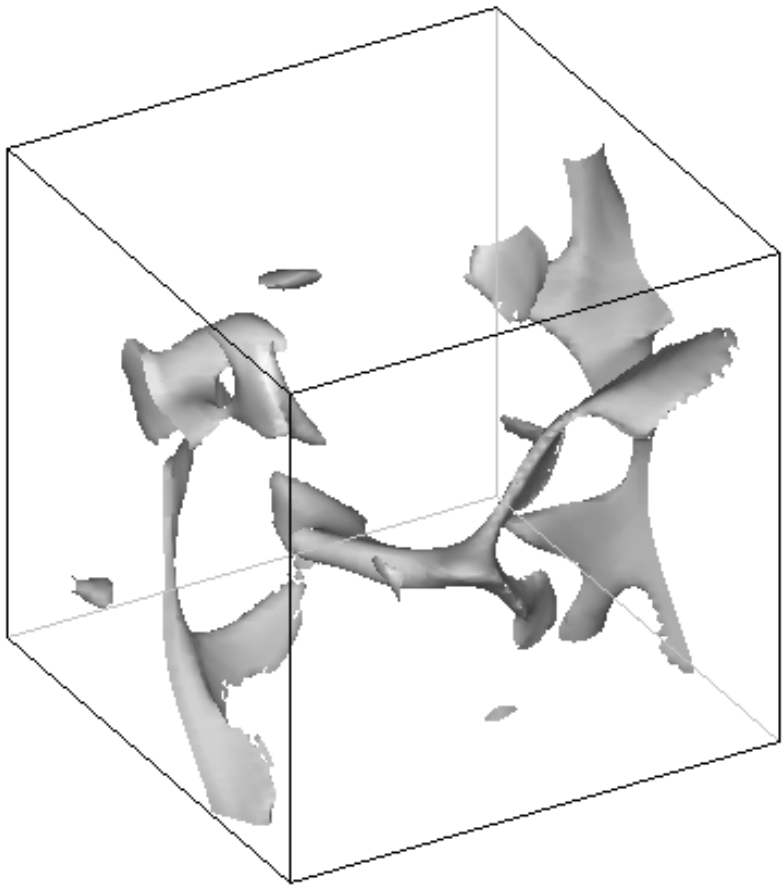

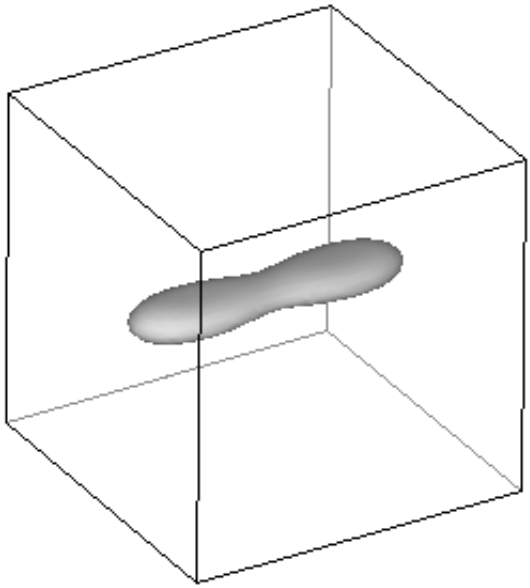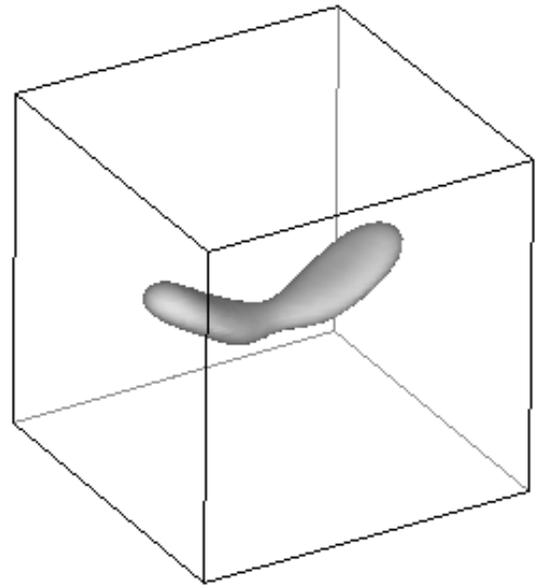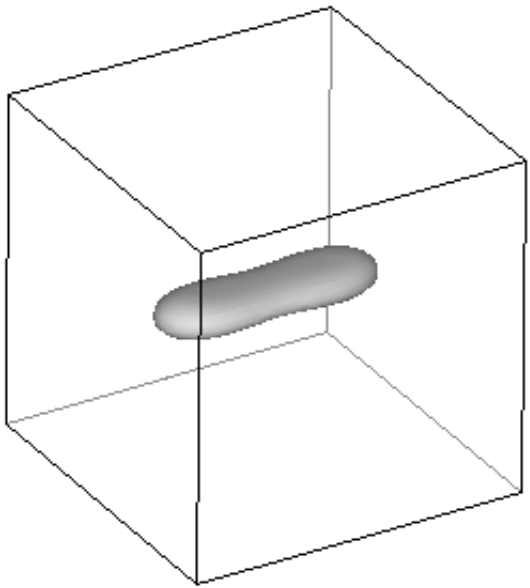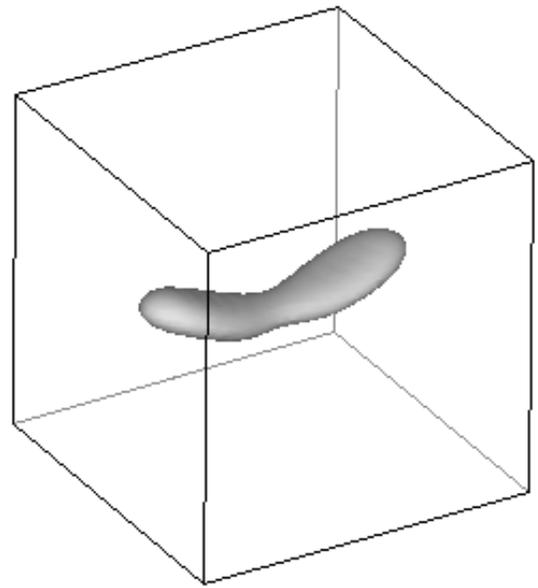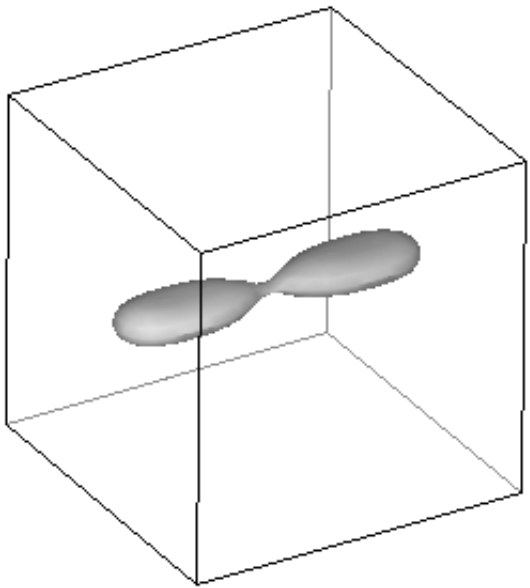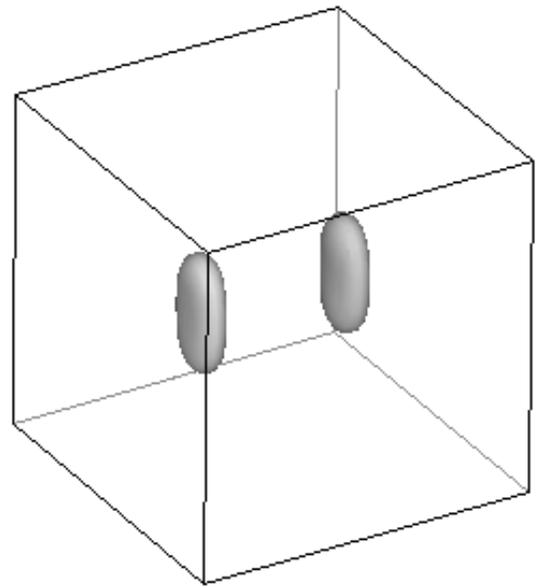

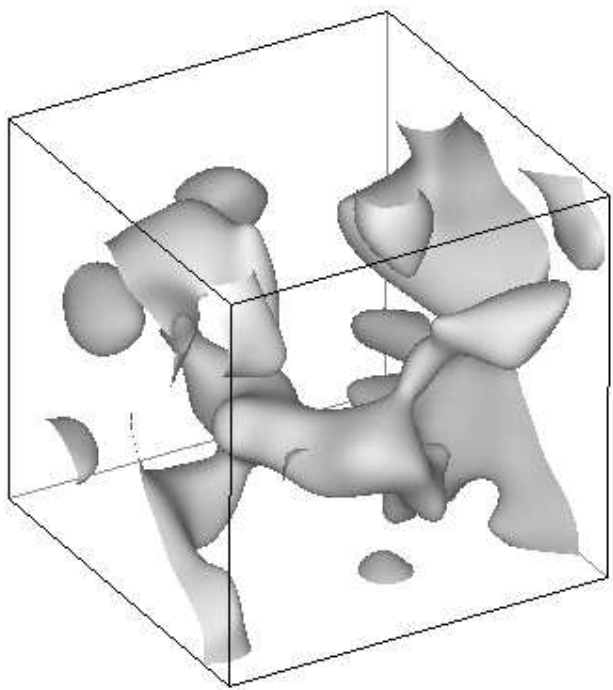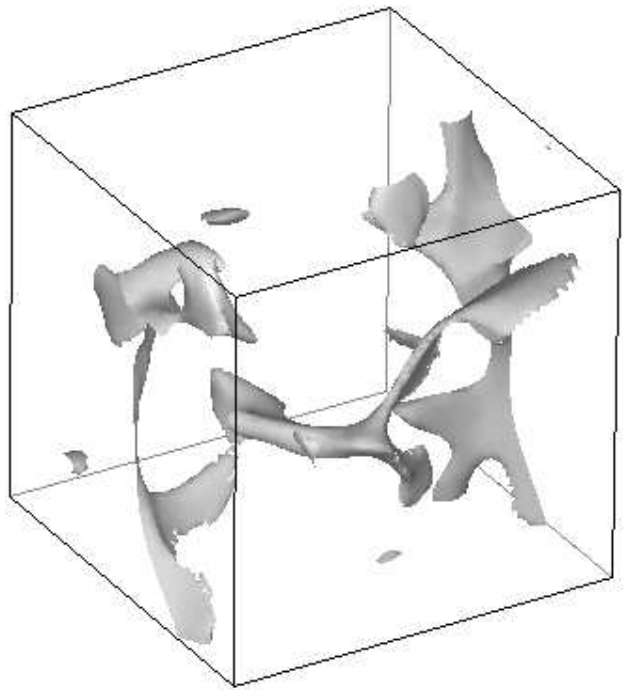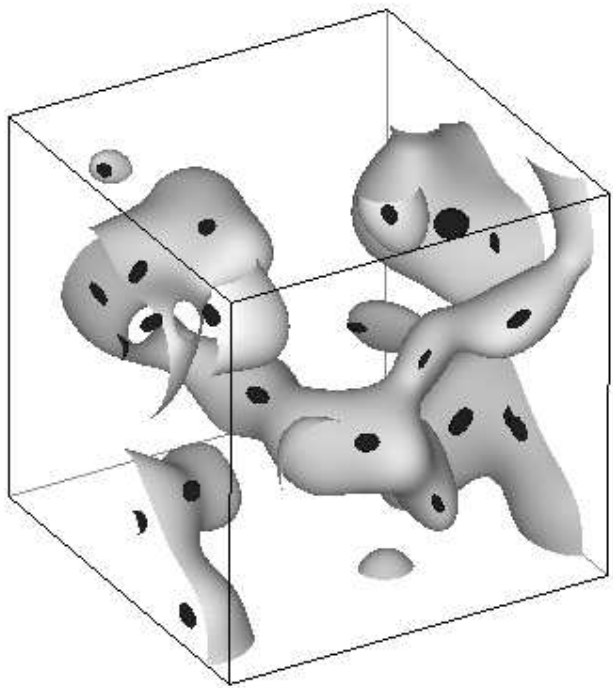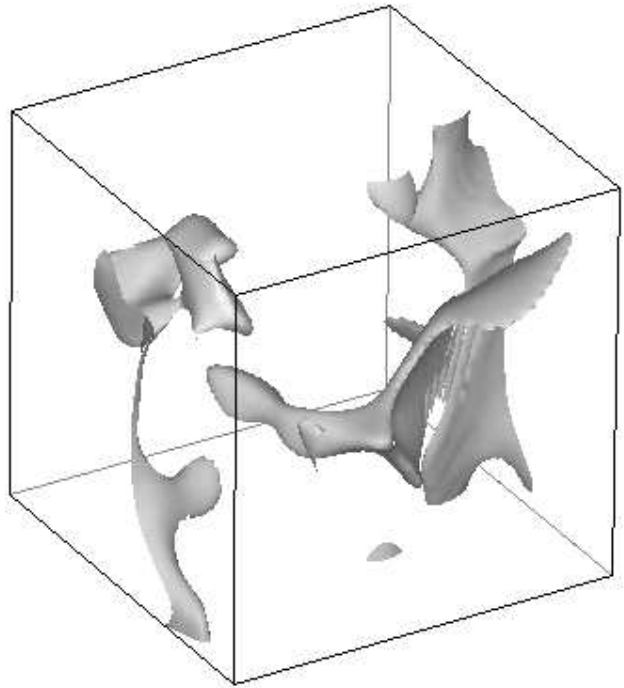

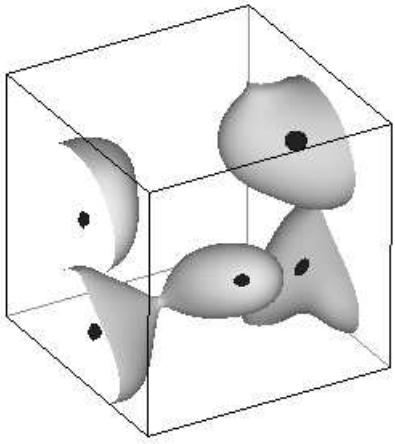
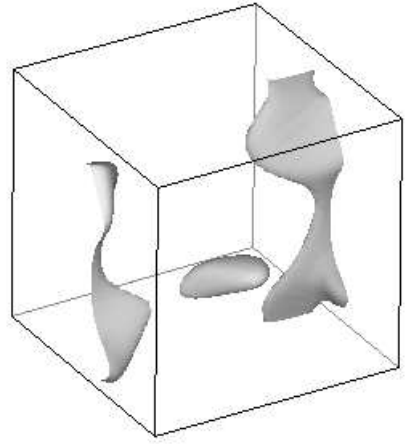
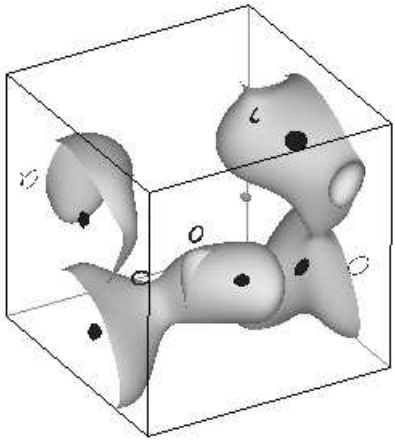
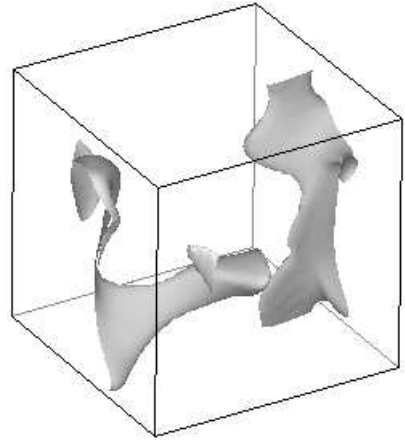
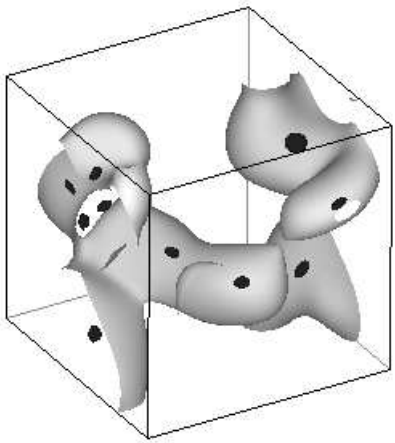
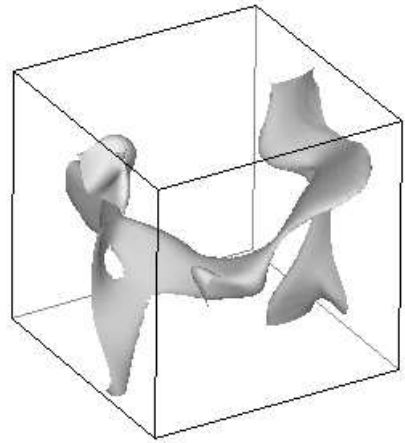
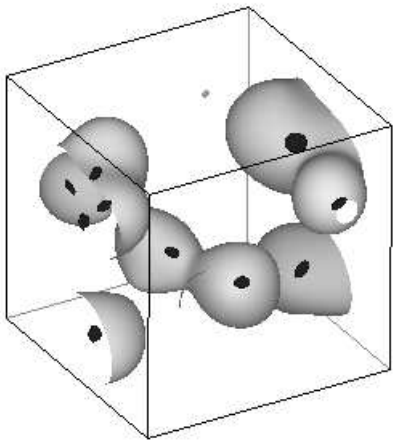
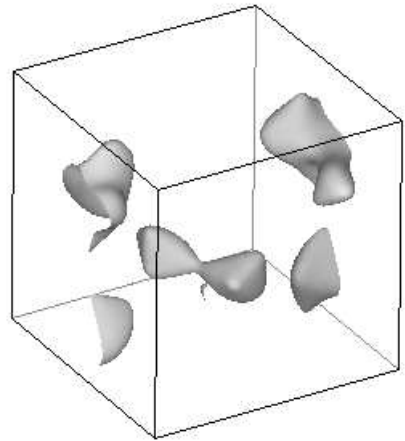

# How Filaments Are Woven Into The Cosmic Web

J. Richard Bond[1], Lev Kofman[2] and Dmitry Pogosyan[1]
1. Canadian Institute for Theoretical Astrophysics
University of Toronto, Toronto ON M5S 1A7, Canada
2. Institute for Astronomy, University of Hawaii, Honolulu HI 96822 USA

Observations indicate galaxies are distributed in a filament-dominated web-like structure. Numerical experiments at high and low redshift of viable structure formation theories also show filament-dominance. We present a simple quantitative explanation of why this is so, showing that the final-state web is actually present in embryonic form in the overdensity pattern of the initial fluctuations, with nonlinear dynamics just sharpening the image. The web is largely defined by the position and primordial tidal fields of rare events in the medium, with the strongest filaments between nearby clusters whose tidal tensors are nearly aligned. Applications of the cosmic web theory to observations include probing cluster-cluster bridges by weak gravitational lensing, X-rays, and the Sunyaev-Zeldovich effect and probing high redshift galaxy-galaxy bridges by low column density Lyman alpha absorption lines.

The large scale structure we observe in the distribution of galaxies is thought to be a consequence of the gravitational instability of a medium that was relatively smooth in the early Universe, with the nonlinear patterns we see having evolved from small density fluctuations superposed on a homogeneous and isotropic background. The basic elements in the pattern are clusters, rich and poor, filaments and sheets, connected in a network, with voids largely devoid of galaxies encompassing the volume in between.[1] $N$-body simulations evolving the best-motivated classes of initial fluctuations into the nonlinear regime also show a network of filaments, a result that was visually evident for many years,[2,3] but not really understood. A sample "cosmic web" $N$-body pattern is shown in Figure 1, for group and cluster formation in a "cold dark matter" model Universe: as the density threshold drops from high values the regions which first emerge are clusters, then arms stretching from the clusters, which ultimately join to form the predominantly filamentary network: the first pattern to percolate is filamentary and it is that which the eye picks out. The basic filamentary pattern is seen on smaller scales at higher redshift in simulations, with bright galaxies playing the role of clusters. There was a suspicion that either the structure was an artifact of simulations or would need fully nonlinear dynamics to understand.[4]

Two classical stories have largely defined how people have thought about structure in the Universe, the (Russian) pancaking picture[5,6] and the (Western) hierarchical clustering picture.[7] The Zeldovich theory[5] applied to models with fluctuation power on cluster scales but none on galactic scales. It led to a highly influential description of the medium, based upon the catastrophe theory of the caustics in the density field[8]: planar pancakes form first, these drain into filaments which in turn drain into clusters, with the entirety forming a cellular network of sheets. The rival hierarchical clustering theory applied to models with initial density fluctuation power decreasing with increasing scale, with a characteristic mass scale for nonlinear objects increasing with time.[9,7] The main paradigm used here was a merging-halos picture, where halos are associated with peaks in the 'background' linear (Gaussian-distributed) density field,[10] *i.e.*, of the rare events in the medium.

The nonlinear dynamical evolution of an initially cold medium is described by a map, $\mathbf{x}(\mathbf{r},t) = \mathbf{r} - \mathbf{s}(\mathbf{r},t)$, from Lagrangian (initial state) space $\mathbf{r}$ to Eulerian (final state) space $\mathbf{x}$. In the early linear phase, the displacement field $\mathbf{s}$ is small and the map is one-to-one, but eventually when particle orbits cross in the medium the map becomes multivalued. We show here that the filamentary web is a consequence of the distribution and spatial coherence of the strain field in the medium. To establish this story, we shall split the displacement field $\mathbf{s} = \mathbf{s}_b + \mathbf{s}_f$ into a background field $\mathbf{s}_b$ smoothed over a large scale $R_b$ and a fluctuating field $\mathbf{s}_f$ built from shorter wavelength components. We also characterize the filtering by the linear *rms* density fluctuation amplitude smoothed on that scale $\sigma_\rho(R_b,t)$. The conceptual picture is one of a smooth stately large-scale single-stream flow $\mathbf{s}_b$ upon which is superposed a relatively chaotic nonlinear short-distance flow $\mathbf{s}_f$ in which orbit crossings abound; we must restrict $\sigma_\rho(R_b,t)$ to be $\lesssim 1$ for this picture to be valid. The linear approximation (in Lagrangian space), $\mathbf{s}_b(\mathbf{r},t) = D(t)\mathbf{s}_b(\mathbf{r})$, where $D(t)$ describes the linear growing mode of fluctuations,[4] defines the "truncated" Zeldovich map, which by itself gives a reasonable description of the overall pattern of $N$-body simulations.

The 'background' strain field is defined by $e_{b,ij}(\mathbf{r}) \equiv -(\nabla_i s_{bj} + \nabla_j s_{bi})/2$. In linear theory, the shear tensor is just its time derivative, $\dot{e}_{b,ij}$, $\mathbf{s}_b$ is proportional to the gradient of the smoothed gravitational potential and hence the smoothed peculiar tidal field acting on a patch of matter is proportional to $e_{b,ij}$. The strain tensor has eigenvalues $-\lambda_{vj}$, which we order by $\lambda_{v3} \geq \lambda_{v2} \geq \lambda_{v1}$. We also express[11] the $\lambda_{vj}$ in terms of the height of the density fluctuation $\nu_b = (\lambda_{v1}+\lambda_{v2}+\lambda_{v3})/\sigma_\rho$, the shear ellipticity $e_v = (\lambda_{v3}-\lambda_{v1})/(2\nu_b\sigma_\rho)$ and prolaticity $p_v = (\lambda_{v1}+\lambda_{v3}-2\lambda_{v2})/(2\nu_b\sigma_\rho)$; thus $e_v \geq 0$ and $-e_v \leq p_v \leq e_v$. In this notation, extreme sheet-like structures would have $p_v \sim e_v$, extreme filaments would have $p_v \sim -e_v$. Both are statistically unlikely: in the conditional distribution[11] $P(\{\lambda_{vj}\}|\nu_b)$, $\langle e_v|\nu_b\rangle \approx 0.54\nu_b^{-1}$ is significantly nonzero while $\langle p_v|\nu_b\rangle$ is zero (and the dispersion is $\sim 0.2\nu_b^{-1}$ for both); and we find[12] regions with $p_v \sim 0$ and large $e_v$ favour filaments over sheets. The conditional distribution $P(\{\lambda_{vEj}\}|\delta_{Zc})$ of the Zeldovich-mapped shear eigenvalues,[13] $\lambda_{vEj} = D(t)\lambda_{vj}[1 - D(t)\lambda_{vj}]^{-1}$, given the Zeldovich-mapped overdensity at the percolation threshold, $\delta_{Zc}$ (where $(1 + \delta_Z) = \prod(1 + \lambda_{vEj})$), is qualitatively similar to $P(\{\lambda_{vj}\}|\nu_b)$, and further favours filaments.[12]

Fig. 1(b) shows the structure of the isodensity surface of the Zeldovich background field map at the percolation threshold. Even with this approximate dynamics, the web defined by filaments that was evident in Fig. 1(a) is clearly seen: sheets are not the dominant elements. In the truncated map with full nonlinear rather than Zeldovich dynamics, the cluster regions occupy less Eulerian volume which leads to an even more enhanced filamentary character than in Fig. 1(b). And the prominent filaments in the final state are there in fattened form in the initial state, as Figure 3 shows.

To quantitatively probe the coherent structures in the initial conditions, we turn to a new language: conditional multiple-point correlation functions in Lagrangian space — *i.e.*, statistically-averaged density and displacement fields subject to various constraints on the shear at multiple points. Consider first the behaviour of an ambient patch about a single random point in the medium at which we specify the shear parameters $e_v$ and $p_v$ to be the typical ones corresponding to a given height $\nu_b$ in the Lagrangian space: we find the typical density contours are more filamentary than sheet-like even before Zeldovich mapping and the density coherence scale along the 1-direction, which is the filamentary axis, is greatly enhanced.

The remarkable size of the filaments is not derivable from the one-point constraint. For this, we turn to correlations constrained by at least two rare peak-patches in the medium. Examples are shown in Figure 2. The peak-patch theory[11] shows that it is statistically likely that, given a specific orientation of the shear tensor for a peak-patch, neighbouring peak-patches will be preferentially along this 1-axis and have shear tensors aligned with it[14]. We allow the separation to vary as well as the shear orientation. What we see is that the strong correlation bridge between two clusters gradually weakens as the separation increases, or as the shear tensor orientations become misaligned. The reason for this phenomenon is that the high degree of constructive interference of the density waves required to make the rare peak-patches, and to preferentially orient them along the 1-axis, leads to a slower decoherence along the 1-axis than along the others,[15] and thus a higher density.

So we paint the following picture how clusters and their filamentary bridges weave a web: Clusters nearer than the mean cluster separation can have strong correlation bridges. There are many such pairs since clusters are statistically biased.[10] The bridge breaks if the separation is too large, isolating the clusters from each other — unless there is a cluster in between to which both have extended their filamentary bridges. To make this into a network, consider laying down the rare events in the medium according to the clustering pattern of peak-patches which become clusters when they evolve dynamically. The correlation bridges arch from cluster to cluster in much of the domain, and these dynamically evolve to filaments, creating the network and containing the bulk of the mass. The typical scale of the segment (bridge) of the filamentary network is $\sim 30 h^{-1}$ Mpc. The order in which the physically significant structures arise is basically the inverse of that in the classical pancake picture: *first high density peaks, then filaments between them, and possibly afterwards the walls, defined as the rest of the mass between voids.*

An important byproduct of this picture is our ability to sequentially capture more and more large-scale structural features by imposing constraints from a rank ordered list of peak-patches. It gives us a method for reconstructing initial states using the rarest peaks (clusters) and, through a background-field map, the main features of the final states. The excellent reconstruction shown in Figure 3 develops as peaks are sequentially added, as is shown in Figure 4. The anisotropic peak-strain constraint is *essential* for success in reconstruction.

The list of peak-patches (and void-patches) is therefore a powerful way to rank order and maximally compress the information stored in the initial conditions, showing what is essential to define structures on the basis of a modest set of local measurements. The characterization is spatially localized: only the rare events nearby define the region. And a modest extrapolation of the work reported here confirms the conjecture of ref.[11]: within a cluster-scale peak-patch, the rarest group-scale peak-patches defined at moderate redshift will define the dominant evolutionary characteristics (the merger history), and one has to work one's way down the group-patch list only if one wants finer and finer dynamical detail. This has important applications to the construction of constrained realizations for nonlinear numerical simulations.

The web pattern depends upon the shape of the initial state power spectrum, $\langle|\delta_L(k)|^2\rangle$, which is usually parameterized by a local power law index $n_{\rho,eff}(k) = d\ln\langle|\delta_L(k)|^2\rangle/d\ln k$. In our figures, we used the standard CDM spectrum, which has $n_{\rho,eff} \sim -1.2$ on cluster scales. When we steepen the spectrum to $n_{\rho,eff} \gtrsim 0$, the clusters become less clustered and the coherence of the web is lost, and although some filaments remain they are weaker and shorter. When we flatten it to $n_{\rho,eff} \lesssim -2$, the clusters are more clustered, the coherence is more pronounced and the filaments are both strengthened and widened. We also see stronger membranes in the regions between the filaments when a number of clusters are close together. These are sheet-like structures, but are not classical pancakes.

There are several immediate applications of our theory to observations. We suggest superclusters are predominantly cluster-cluster bridges; no image from the cosmology of the 80's was as powerful as the CfA picture[16] of the Coma cluster and its environs, with a great arm of galaxies spanning Coma and Abell 1367. "Great Wall" features would be interpreted as the membrane-like webbing in between the filaments, and would be more pronounced the flatter the spectrum; e.g., theories which match cluster observations have $n_{\rho,eff} \sim -1.6$ on cluster scales, hence on average would be more membrane-like than CDM would be.

We predict that the strongest filaments exist between highly clustered and aligned clusters. This should be especially notable in the mass distribution around systems of clusters reconstructed with weak gravitational lensing,[17] which uses the projected strain (shear) field. It also may affect the strategy of the search for X-ray gas[18] or the Sunyaev-Zeldovich effect in filaments. Most theories predict density fluctuation spectra are flatter at galactic than cluster scales, with $n_{\rho,eff} \lesssim -2$ appropriate for redshifts $z \gtrsim 3$. The neutral hydrogen absorption by the filaments existing then would be observed as part of the quasar Lyman alpha forest,[19] with column densities $N_{HI} \lesssim 10^{14.5}\text{cm}^{-2}$ and a characteristic (comoving) filamentary size roughly the distance between dwarf galaxies, $\sim 1\,h^{-1}\text{Mpc}$. As well, we expect filaments of dwarf galaxies between normal galaxies at similar redshifts.

ACKNOWLEDGEMENTS: We thank S. Myers for discussions. Support was provided by NSERC, the Canadian Institute for Advanced Research and the IFA.

**Figure Captions**

**Figure 1.** (a) $N$-body simulation (courtesy of A. Klypin) of a cold dark matter (CDM) model with $128^3$ particles in a comoving periodic boundary volume of $(50 \, h^{-1} \mathrm{Mpc})^3$ evolved to the *rms* density fluctuation amplitude $\sigma_8 = 0.7$ (where $\sigma_8 = \sigma_\rho$ at a top-hat filtering scale $R_b = 8 \, h^{-1} \mathrm{Mpc}$, where h parameterizes the Hubble constant, $H_0 = 100 h \, \mathrm{km \, s^{-1} \, Mpc^{-1}}$). 1 in 20 particles are plotted. The pattern of connected filaments joining clusters evident here is quite general for viable hierarchical theories that explain the galaxy clustering data.

(b) Truncated Zeldovich map for the (Lagrangian space) initial conditions that led to the Eulerian space pattern of (a). The initial state was smoothed on a Gaussian scale of $R_b = 5 \, h^{-1} \mathrm{Mpc}$, corresponding to a linear amplitude of $\sigma_\rho = 0.54$. The density contour level shown, $\delta_{Zc} = 1$, corresponds to the percolation threshold. At high threshold ($\delta_Z \gtrsim 4$), the contours are isolated around rare massive clusters, at medium threshold ($\delta_Z \sim 2$), arms radiate from the clusters, and there are some web-like membranes close to the clusters; by $\delta_{Zc} = 1$, the filaments connect the clusters. Making $R_b$ smaller shows more detailed structure, and the filaments can break up into beads upon them (the eye still catches the coherence), but the picture is otherwise similar. As we see here, the Zeldovich map does not imply that pancakes associated with caustics are the dominant overdensity enhancement structures in the medium for $\sigma_\rho(R_b) = \mathcal{O}(1/2)$. (We have also checked whether the contours of $\lambda_{vE3}$ bear much relationship to the observed structures, as might be expected in the classical pancake scenario: they do not.) It is the rare positive and negative peaks that condition the overall pattern formation in the medium. We identify the filaments that are so visually evident as "correlation bridges" between the collapsed peak-patches, forming a web, but no classical pancakes.

**Figure 2.** Zeldovich-mapped correlation functions constrained by shear at two peaks show the effect of changing separation on the correlation bridge and changing shear-tensor orientation. What we see is that the strong correlation bridge between two clusters gradually weakens as the separation increases, breaking by $50 \, h^{-1} \mathrm{Mpc}$, or as the shear tensor orientations become misaligned. For the left panels, shear tensors are aligned, and the separations are $40 \, h^{-1} \mathrm{Mpc}$ (top), $30 \, h^{-1} \mathrm{Mpc}$ (middle) and $50 \, h^{-1} \mathrm{Mpc}$ (bottom, with the broken bridge). For the right panels, the separation is $40 \, h^{-1} \mathrm{Mpc}$ and the misalignment angles relative to the line joining the peaks are: $\pm 30°$ (top), $\pm 20°$ (middle) and $\pm 90°$ (bottom), when the bridge is fully broken. (At this $\delta_{Zc} = 1$ level with $\pm 45°$ orientation the bridge is also broken, but joins at a lower threshold.) Mean-shear parameters appropriate for Abell-cluster peak-patches were adopted: $\nu = 3$, $e_v = 0.18$ and $p_v = 0$. (The average separation of clusters of this richness is $\approx 50 \, h^{-1} \mathrm{Mpc}$ and their correlation length is $\sim 20 \, h^{-1} \mathrm{Mpc}$; *i.e.*, clusters come in patches.[15]) To avoid too much interior dynamics in the patches, the truncated Zeldovich map was used with Gaussian smoothing scale $R_b = 5 \, h^{-1} \mathrm{Mpc}$ ($\sigma_\rho = 0.54$), the same as for the peaks (whose top hat scale is $R_{pk} \approx 10 \, h^{-1} \mathrm{Mpc}$). Note that in Lagrangian space the length of the filament bridging the two peak regions is similar to the diameter of the peak patches. The contour shown is at $\delta_{Zc} = 1$, about where percolation occurs in the $N$-body computation at this $\sigma_\rho$ smoothing.

**Figure 3.** The upper left panel shows the linear CDM density field $\delta_L(\mathbf{r})$ in Lagrangian space for the $(50 \, h^{-1} \mathrm{Mpc})^3$ box of Fig. 1. It has been smoothed on a Gaussian scale $R_b = 3.5 \, h^{-1} \mathrm{Mpc}$, corresponding to $\sigma_\rho = 0.65$. The threshold chosen is $\delta_L = 1\sigma_\rho$ because the linear density field percolates at this one-sigma level. The upper right panel, a repeat of Fig. 1(b), is the Zeldovich-

map of the smoothed initial conditions into Eulerian space, for a contour threshold $\delta_Z = 2$, just above where percolation occurs. The final state filaments are present in the initial state. The lower left panel shows the linear density correlation function $\langle \delta_L(\mathbf{r})|20\,\mathrm{pks}\rangle$ constrained by the top 20 peak-patches identified in the initial conditions, and the lower right shows its Zeldovich map.

This mean field reconstruction of $\delta_L$ given "observations" of the values of the shear tensor $e_b^{ij}$ and the displacement $\mathbf{s}_b$ at $N$ peak-patch points $\mathbf{r}_{pk}$ averaged over the peak-patch size $R_{pk}$, $\langle \delta_L|N\,\mathrm{pks}\rangle$. This is a straightforward application of a well-known theorem for a constrained Gaussian field, ref.[7], Appendix D. This gives $9N$ constraints. The requirement that the gradient of the density field vanishes at $\mathbf{r}_{pk}$ (the extremum constraint) adds a further $3N$ constraints, but is unimportant in practice. Although the positions are not counted explicitly as constraints, changing them radically alters the way the mean fields look. The greater $N$ is, the smaller are the fluctuations about the mean.

The patches are indicated by blackened ellipsoids of overall size proportional to $R_{pk}$ and shape defined by the shear tensor orientation, with the shortest axis corresponding to the highest shear eigenvalue, the longest to the least. [To find peak-patches[11], we first identify peaks in the smoothed linear density field above a threshold $\delta_{Lc} = 1.686$ (the linear overdensity needed for a spherically symmetric "top hat" to collapse to a point); the smoothing is done on a hierarchy of Gaussian filter scales, here, 25 filters ranging from 1 to $4\,\mathrm{h}^{-1}\mathrm{Mpc}$. This of course greatly overcounts sites of local collapses — a very strong version of the "clouds-in-cloud" problem that is present even for single-filter peaks, as noted in ref.[10]. About each of these points, we calculate the radius $R_{pk}$ at which the interior region is predicted to have just collapsed in all three dimensions at the redshift in question. At a higher redshift, the radius would be smaller, and at a lower redshift larger. For this we use the homogeneous ellipsoid model[11] for interior dynamics which utilizes both the interior tidal field and a linear-evolution approximation to the exterior tidal field acting on the patch. This requires measuring the strain (or shear) tensor within this radius (top-hat filtered). The technique[11] also gives the local displacement field averaged over the Lagrangian volume within. In the next stage of the peak-patch procedure, smaller scale peaks whose centres lie within larger scale peaks are removed from the list. The net list is rank-ordered in mass. We find about 400 peak-patches in our $(50\,\mathrm{h}^{-1}\mathrm{Mpc})^3$ volume, with the top 20 ranging in size $R_{pk}$ from 4.5 to $9.2\,\mathrm{h}^{-1}\mathrm{Mpc}$, with masses from $10^{14.3}$ to $10^{15.3}\,\mathrm{M}_\odot$.]

**Figure 4.** The top three panels show mean fields with an increasing number of peak-patch and void-patch constraints imposed. Left shows Lagrangian space, right shows the result after Zeldovich-mapping into Eulerian space. The top panels use the top 5 peaks, the second use the top 5 voids (treated as "inverse-peaks") as well as the top 5 peaks, and the third use the top 10 peaks. Adding 5 voids does not do as well as adding 5 more peaks. The fourth set, using constraints on the peak height and the velocity (or displacement field) only, shows how important the anisotropic shear constraints are for defining the structure. The peaks used are shown as solid ellipsoids, the voids as open ellipsoids, scaled in size $\propto R_{pk}$, with shape defined by the shear tensor, shortest axis corresponding to the highest shear eigenvalue, longest to the least. At the thresholds $\delta_L = 1$, $\delta_Z = 2$ of Fig. 3 some strong bridges seen in the 20 peak reconstruction, $\langle \delta_L|20\,\mathrm{pks}\rangle$, are not as evident in the $\langle \delta_L|10\,\mathrm{pks}\rangle$–field, but emerge at lower thresholds; $\delta_L = 0.5$ and $\delta_Z = 1.3$ are shown here. This figure shows that it is better to use the top $2N$ peaks rather than $N$ voids and $N$ peaks if we are focusing only on regions of higher density contrast — which is observationally most relevant for most astrophysical observables. The situation would be reversed if it were voids we were concentrating on. It is rather interesting to see how the void-patch constraints create high mean field regions in between them, just where less rare peak-patches reside. It is just not as precise a way to get that structure for the same computational effort.